\documentclass[12pt,preprint]{aastex}
\usepackage{epsfig}
\usepackage{natbib}

\begin{document}
\voffset-0.5cm
\newcommand{\gsim}{\hbox{\rlap{$^>$}$_\sim$}}
\newcommand{\lsim}{\hbox{\rlap{$^<$}$_\sim$}}

\title{GRB 130603B: No Compelling Evidence For Neutron Star Merger}

\author{Shlomo Dado\altaffilmark{1} and Arnon Dar\altaffilmark{1}}

\altaffiltext{1}{Physics Department, Technion, Haifa 32000, Israel}

\begin{abstract} 

Near infrared (NIR) flare/rebrightening in the afterglow of the short hard 
gamma ray burst (SHB) 130603B measured with the Hubble Space Telescope 
(HST) and an alleged late-time X-ray excess were interpreted as possible 
evidence of a neutron-star merger origin of this SHB. However, the X-ray 
afterglow that was measured with the Swift-XRT and Newton XMM have the 
canonical behaviour of a synchrotron afterglow produced by a highly 
relativistic jet. The H-band flux observed with HST 9.41 days after burst 
is that expected from the measured late-time X-ray afterglow. A late-time 
flare/re-brightening of a NIR-Optical afterglow of SHB can be produced by 
jet collision with an interstellar density bump, or by a kilonova, but jet 
plus kilonova can be produced also by the collapse of compact stars 
(neutron star, strange star, or quark star) to a more compact object due 
to cooling, loss of angular momentum, or mass accretion.

\end{abstract}


\maketitle

\section{Introduction}

Stripped envelope supernova explosions and neutron star mergers in close 
binaries were originally suggested by Goodman, Dar and Nussinov~(1987) as 
possible sources of cosmological gamma ray bursts. However, their proposed 
underlying mechanism - a spherical fireball produced by 
neutrino-antineutrino annihilation into electron positron pairs beyond the 
surface of the collapsing/merging star- turned out not to be powerful 
enough to produce GRBs observable at very large cosmological distances.  
Consequently, Shaviv and Dar~(1995) proposed that highly relativistic jets 
of ordinary matter are probably ejected in such events, and produce 
narrowly collimated GRBs by inverse Compton scattering of circumstellar 
light. They also suggested that short GRBs may also be produced by highly 
relativistic jets ejected in the phase transition of compact stars, such 
as neutron stars, strange stars and quark stars, into more compact objects 
due to mass accretion or to cooling and loss of angular momentum via winds 
and radiation. After the discovery of GRB afterglows, Dar (1998) proposed 
that they are highly beamed synchrotron radiation produced in the 
collision of these highly relativistic jets with the interstellar matter.

By now, there is convincing evidence that long duration GRBs are produced 
mostly by highly relativistic jets launched in stripped envelope supernova 
explosions (mainly of type Ic), but, despite enormous observational 
efforts, the origin of short duration GRBs remains unknown. In fact, the 
circumstantial evidence that has been claimed to link short hard GRBs 
(SHBs) with neutron star mergers in close binaries, such as their location 
in both spiral and elliptical galaxies and the distribution of their 
location offsets relative to the center of their host galaxies, which 
extend to a distance of ~100 kpc (e.g., Berger et al.~2013 and references 
therein) and beyond? (e.g., SHB 080503 with the lack of a coincident host 
galaxy down to 28.5 mag in deep Hubble Space Telescope imaging, Perley et 
al.~2009)  actually favours single compact stars with large natal kick 
velocity as progenitors of SHBs (Dado, Dar and De R\'ujula 2009a) over 
neutron star binaries, which have much lower Galactic velocities.

It was predicted that neutron star mergers create significant quantities of 
neutron-rich radioactive nuclei whose decay should result in a faint 
transient in the days following the burst, a so-called kilonova or 
macronova (Li and Paczynski 1998). Recently, the broad band afterglow of 
the SHB 130603B (Melandri et al.~2013; Golenetskii et al.~2013) 
that was measured with the Swift X-ray telescope 
(XRT), Newton XMM, HST and ground based optical and radio telescopes was 
interpreted as possible evidence of a neutron-star merger origin of SHB 
130603B (Tanvir et al.~2013a,b; Berger et 
al.~2013; Fong et al.~2013). However, in this letter we show that the 
X-ray afterglow of SHB 130603B, which was measured with Swift XRT 
(Swift/XRT GRB light-curve repository, Evans et al.~2009) and Newton XMM 
(Fong et al.~2013) has the canonical behaviour of a synchrotron afterglow 
produced by a highly relativistic jet propagating in a normal interstellar 
environment, as predicted by the cannonball model of GRBs (Dado, Dar and 
De R\'ujula 2002, 2009a,b; Dado and Dar~2013) long before its empirical 
discovery (Nousek et al.~2006). This canonical X-ray afterglow does not 
have a "mysterious late-time X-ray excess" as claimed in Fong et al.~2013, 
and the flux observed in the NIR H-band with HST 9.6 days after burst 
(Tanvir et al.~2013a,b) is that expected from the measured late-time X-ray 
afterglow.

Moreover, a fast decline of a late-time afterglow followed by a 
re-brightening/flare in the NIR and optical afterglow of a GRB can be 
produced by a jet colliding with a density bump in the interstellar medium 
(e.g., Dado, Dar and De R\'ujula 2002, 2009b), as was observed in several 
long-duration GRBs, such as 030329 (Lipkin et al. 2004; Matheson et 
al.~2003) 070311  (Guidorzi et al.~2007) 
and SHBs such as 050724 (Malesani et al.~2007) and 080503 
(Perley et al.~2009). The host galaxy of SHB 130603B at redshift ${\rm 
z=0.356}$ (Thone et al.~2013), as seen in high-resolution HST imaging 
(Tanvir et al.~2013b), 
is a perturbed spiral galaxy due to interaction with another galaxy. SHB 
130603B was located  in a tidally disrupted 
arm. The interaction of the SHB jet with such a bumpy environment may have 
caused the flare/re-brightening in the NIR afterglow observed with the HST 
on day 9.41.

Furthermore, a late-time flare/re-brightening of a NIR-Optical afterglow 
of SHB can be produced by either a jet collision with an interstellar 
density bump or a kilonova, but jet plus kilonova can be produced also by 
collapse 
of compact stars (neutron star, strange star, or quark star) to a more 
compact object due to cooling, loss of angular momentum, or mass 
accretion. In fact, single compact stars are more likely to be found at 
large offsets from galactic center/disks (e.g., in the galactic halos or 
the intergalactic medium) than neutron star binaries or black hole-neutron 
star binaries.

\section{The X-ray afterglow of  SHB 130603B} 

The conclusion of Fong et al.~(2013)  that the X-ray  afterglow of SHB 
130603B 
shows "a mysterious late-time X-ray excess" was  
based on a standard fireball model analysis 
of its  X-ray afterglow.  
The standard fireball model, however, predicts that the temporal index 
$\alpha$ of the afterglow of a conical jet that is 
parametrized as a smoothly broken power-law, 
${\rm  F_{\nu}\propto t^{-\alpha}\, \nu^{-\beta}}$, 
increases by
${\Delta\alpha_X = 0.75}$ across the jet break, independent of
the spectral index ${\rm \beta_X}$ of the 
afterglow (Dado and Dar 2013).
The temporal indices ${\rm \alpha_X=0.35\pm0.08}$
and ${\rm \alpha_X=1.61\pm 0.08}$ 
before and after the break, respectively,
that were reported for instance in the Swift-XRT GRB Catalogue (Evans et 
al.~2009) yield ${\Delta\alpha_X = 1.26 \pm 0.11}$ 
(${\rm \alpha_X=1.75\pm 0.08}$
for the combined data of Swift-XRT and Newton XMM  
yield ${\Delta\alpha_X = 1.40 \pm 0.11}$), which is at odds 
with the conical fireball model. 
   
In contrast, the  X-ray afterglow of SHB 130603B that was  measured 
with Swift-XRT (Swift/XRT light curve repository, Evans et al.~2009)
and Newton XMM (Fong et al.~2013)  
has the canonical behaviour of a normal synchrotron afterglow 
produced by a highly relativistic jet propagating in a normal interstellar
environment of its host galaxy (Dado, Dar and De R\'ujula 2009a,b). 
This canonical behaviour of the X-ray afterglow was predicted  
(Dado, Dar and De R\'ujula~2002) long before the launch of Swift and 
its empirical discovery (Nousek et al.~2006). It consists of an 
early plateau 
phase that follows the fast decline phase of the prompt emission and
breaks smoothly  into a late-time  (${\rm t\gg t_b}$)
power-law decline with a power law index that 
satisfies the cannonball (CB) model
closure relation, 
\begin{equation}
{\rm \alpha_X=\beta_X+1/2=\Gamma_X-1/2}\, ,
\label{Closure}
\end{equation}
independent of the pre-break power-law index,
where ${\rm \Gamma_X}$
is the photon spectral index of the X-ray
afterglow (see, e.g., Dado and Dar 2013 and references therein).
Using the value ${\beta_X=1.15\pm 0.11}$, which was obtained by 
de Ugarte Postigo et al.~(2013) from the Swift-XRT data, 
the CB model closure relation yields ${\rm \alpha_X=1.65 \pm 
0.11}$. This value is consistent within errors with the post break value 
${\rm \alpha_X=1.61\pm 0.08}$ reported for SHB 130603B
in the Swift-XRT GRB Catalogue (Evans et al.~2009).

In the CB model, the canonical light-curve of the X-ray afterglow 
depends 
only on three parameters (Dado, Dar and De R\'ujula 2009a): the product 
${\rm \gamma\, \theta}$ of the bulk 
motion Lorentz factor of the jet and the viewing angle relative to the 
direction of motion of the jet, the jet deceleration parameter ${\rm 
t_0}$, and the spectral index ${\rm p_e}$ of the Fermi accelerated 
electrons in the jet that satisfies ${\rm p_e=2\,\beta_X}$.  A CB
model fit to 
the light-curve of the 0.3-10 keV X-ray afterglow of SHB 130603B, which 
was measured with Swift XRT (Evans et al. 2009)  and with Newton XMM
assuming the spectral index that was measured by Swift, is 
shown in Fig.~1. The best fit value ${\rm p_e=2.37}$ yields ${\rm 
\beta_X=p_e/2=1.18}$, which is consistent with the late-time photon index 
${\rm \Gamma_X=2.21\pm 0.18}$ that is reported in the Swift-XRT GRB 
Catalogue (Evans et al. 2009). The two other best fit parameters, ${\rm 
\gamma\, \theta=0.55}$ and ${\rm t_0=878\, s}$, yield a 
deceleration break (so called "jet break") at ${\rm t_b\approx 1500\, s}$.

Thus, we conclude that there is no evidence for a "mysterious late-time 
X-ray excess" that was claimed in Fong et al.~2013,  
and was  explained by a magnetar contribution to the 
afterglow emission of SHB 130603B (Fong et 
al.~2013; Metzger and Piro 2013; Fan et al. 2013).

\section{The near infrared-optical afterglow}

The conclusion that the NIR-optical afterglow of SHB 130603B provides  
possible evidence of a macronova/kilonova was based on a re-brightening 
of the NIR afterglow observed with the Hubble space telescope (HST) in the 
H band on day 9.41, which is well above that extrapolated from the fast 
decline of the optical afterglow in the r band during the first day after 
the break around 0.3 d (Berger and Fong~2013, Berger 
et al.~2013, Fong et al.~2013).

However, in the CB model, when the spectral index of the late-time 
NIR and optical bands is  above the spectral break, 
${\rm \beta_H\approx\beta_X}$ 
and consequently ${\rm \alpha_H=\beta_H+1/2 \approx \alpha_X}$.
Using the ground-based JK-band observations
extrapolated to the H band (Fong et al.~2013)
and the HST H-band measurement,  we obtained that
${\rm \alpha_H=1.61\pm 0.08}$ in the time interval 0.61-9.41 d,
which is in agreement, within errors, with the power-law index  
${\rm \alpha_X=1.68 \pm 0.08}$ of the joint late-time Swift-XRT 
observations 
(Evans et al.~2009) and XMM Newton observations (Fong et al.~2013).

Moreover, a broken power-law best fit to the unabsorbed late-time broad 
band NIR, Optical and Swift X-ray spectrum by de Ugarte Postigo et al. 
(2013) yielded ${\beta=0.65\pm 0.09}$ below a break at ${\rm 
\nu_b=9.55\times 10^{15}\,Hz}$ and ${\beta_X=1.15\pm 0.11}$. Using 
${\lambda=12.4}$A for 1 keV photons, ${\lambda=16300}$A for H-band 
photons and ${\rm \lambda_{break}=314}$A, the expect flux ratio of the H 
and X-ray bands is ${\rm F_H/F_{keV}\approx 536 \pm 160}$. This ratio is 
in good agreement within errors with the observed ratio ${\rm 
F_H/F_{keV}=623\pm 160}$ of the H-band flux measured with HST on day 9.41 
after burst and the 1 keV X-ray flux obtained by extrapolating the 
joint Swift/XRT - Newton XMM 1 keV flux to day 9.41.

The highly relativistic jets of plasmoids (cannonballs) that produce GRBs 
can encounter a bumpy interstellar medium in the host galaxy. Also, the 
opacity along the line of sight to the jet in the host can vary 
significantly due to the "superluminal" motion of the line of sight 
to the jet in the 
host galaxy. The collision of a jet with an over-density bump can produce 
chromatic re-brightening/flare in the NIR-Optical afterglow (e.g., Dado, 
Dar and De R\'ujula 2002, 2009b)  as was observed in the late-time optical 
afterglow of several long duration GRBs such as 030329 (Lipkin et al. 
2004; Matheson et al.~2003) and SHBs such as 050724 (Malesani et 
al.~2007) and 080503 (Perley et al.~2009), while under density can 
cause a fast temporal decline of an afterglow (${\rm \alpha >2}$), as 
observed in several GRBs (Swift/XRT GRB catalouge, Evans et al.~2009).
Such density variations cause spectral  
and temporal variations in the afterglow, which otherwise has a smooth 
power-law behaviour. After an over density or an under density, the 
late-time 
(${\rm t\gg t_b}$) closure relation of the CB model is recovered when the 
column density as function of distance converges to that of a the mean ISM 
density.  This can explain both a fast decline of the NIR-optical 
afterglow of SHB 130603B after an over density followed by an under 
density, and a recovery to the normal power-law 
decline like that of the X-ray afterglow.

\section{The Macronova - SHB association}

SHBs may be produced by highly relativistic jets launched 
in the collapse of compact stars
(neutron star, strange star, or quark star) to a more compact object due 
to loss of angular momentum, cooling, or mass accretion (Dar et al.~1992).
During neutron star merger, or collapse of a compact star 
to a more compact object, the crust layers can 
be stripped off by very strong outgoing shocks. Neutrino-antineutrino 
annihilation into electron-positron pairs behind such blown off  layers 
can than produce fireball with a super Eddington luminosity (Goodman, Dar and 
Nussinov 1987), which can accelerate the blown off crust layers  to 
velocities well 
above the escape velocity (Paczynski 1990; Dar et al.~1992),
althogh the neutrino luminosity, which is well below the 'neutrino 
Eddington luminosity', by itself cannot 
blow off the crust layers: 
  
Balancing the gravitational force with the rate of momentum 
deposition by neutrinos, and neglecting general relativistic effects, 
Dar et al. (1992) derived a 'neutrino Eddington luminosity'
of a compact star 
\begin{equation}  
{\rm L_E(\nu)\approx {4\, \pi\ G\, M\, m_n\, c\over \bar\sigma}\,
\approx 3.74\times 10^{54}\, {M\over M_\odot}\, \left[{E_\nu \over 15\, 
MeV}\right]^2\,  erg}  
\label{ENL}
\end{equation} 
where G is Newton's gravity constant, M is the mass of the compact star, 
${\rm m_n}$ is the mass of a nucleon, and ${\rm \bar{\sigma}\approx 
10^{-43}\, (E_\nu/MeV)^2\, cm^2}$ is the averaged cross section for 
momentum transfer to a nucleon by charged- and neutral-current 
scatterings. The gravitational binding energy release in the stellar 
collapse is transported by neutrino diffusion to the neutrino sphere 
from 
where it escapes as a black body neutrino emission with a typical 
temperature of a few MeV. Such an emission from a neutrino sphere of a 
compact star of radius ${\rm R \sim 10\, km}$ lasts a few ten seconds or 
more, during which the neutrino luminosity is well below the neutrino 
Eddington luminosity of the newly formed compact star.

It is unclear whether a robust r-process occurs in the ejecta, or 
whether neutrinos drive the composition towards $^{56}$Ni dominated 
composition (e.g., Surman et al.~2008). All together, the total mass of 
the ejecta, its composition, density, and velocity, and their radial and 
angular distributions are highly uncertain, which makes the predicted 
signal from an associated macronova (Li and Paczynski 1998)  very 
uncertain and unreliable for distinguishing between 
a single compact star and a binary compact star origin of SHBs.

\section{Conclusions}
 
Several explanations of the re-brightening of the NIR afterglow of SHB 
130603B around 9.41 days after burst have been proposed. These include a 
macronova/kilonova produced by a neutron star merger in a close binary due 
to gravitational wave emission (Tanvir et al.~2013b; Fong et al.~2013), an 
active millisecond magnetar produced in a neutron star merger in close 
binaries (Metzger and Piro 2013), and a late-time flare 
produced by collision of the SHB jet with an ISM density bump
(Dado, Dar and De R\'ujula 2002, 2009a).

A jet plus a mini-supernova/macronova/kilonova, however, are not unique to 
the neutron star merger scenario. They can be produced also in a 
phase-transition/collapse of compact stars (neutron star, strange star or 
quark star) to a more compact object due to cooling, loss of angular 
momentum or mass accretion.

The X-ray afterglow that
was measured with the Swift-XRT and Newton XMM has the expected canonical
behaviour of a synchrotron afterglow produced by a highly relativistic
jet. Its late time behaviour does not support a milisecond magnetar as 
the power-source of the chromatic afterglow of SHB 130603B. 
The late-time H-band flux observed with HST 9.41 days after burst 
is that expected from an ordinary synchrotron radiation from a jet that 
produced the measured late-time X-ray afterglow. 

Late-time flare/re-brightening of a NIR-Optical afterglow of an SHB can be 
produced also by jet collision with an interstellar density bump,
as seen in several GRBs. The host galaxy of SHB 130603B as seen in
high-resolution HST imaging (Tanvir et al.~2013b) is a perturbed spiral
galaxy due to interaction with another galaxy. The GRB was located in
in a tidally disrupted arm of its host galaxy. The interaction of
the GRB jet with such a bumpy environment could produce the
flare/re-brightening of the NIR afterglow of the GRB observed with the HST
9.41 days after burst.

The star formation within the host, location of SHB 130603B on top of the 
tidally disrupted arm, strong absorption features and large line of sight 
extinction that were observed indicate that the GRB progenitor was 
probably not far from its birth place (de Ugarte Postigo et al.~2013), 
untypical to the usually long life time before neutron star merger due to 
gravitational wave emission in the known neutron star binaries in our 
galaxy. Moreover, the failure to detect a host galaxy 
down to 28.5 mag in deep Hubble Space Telescope imaging
searches in the case of e.g., SHB 080503 with a late time 
flare/rebrightening (Perley et al.~2009) suggest a large natal kick 
velocity of its progenitor, unlikely for compact binaries, but often 
observed for isolated neutron stars/pulsars  (Hobbs et al.~2005).

The true smoking gun for the neutron star merger in close binaries is the 
detection of gravitational waves. Unfortunately, this is unlikely to occur 
before the completion of the new generation of gravity-wave detectors, as 
the sensitivity of current detectors such as LIGO, and Virgo, is several 
orders of magnitude below what would be required to detect a merger at a 
distance similar to the nearest SHBs with known redshift.

{\bf Acknowledgement: We thank  A. de Ugarte Postigo for useful comments.}

\newpage 
\begin{figure}[]
\centering 
\vspace{-2cm} 
\epsfig{file=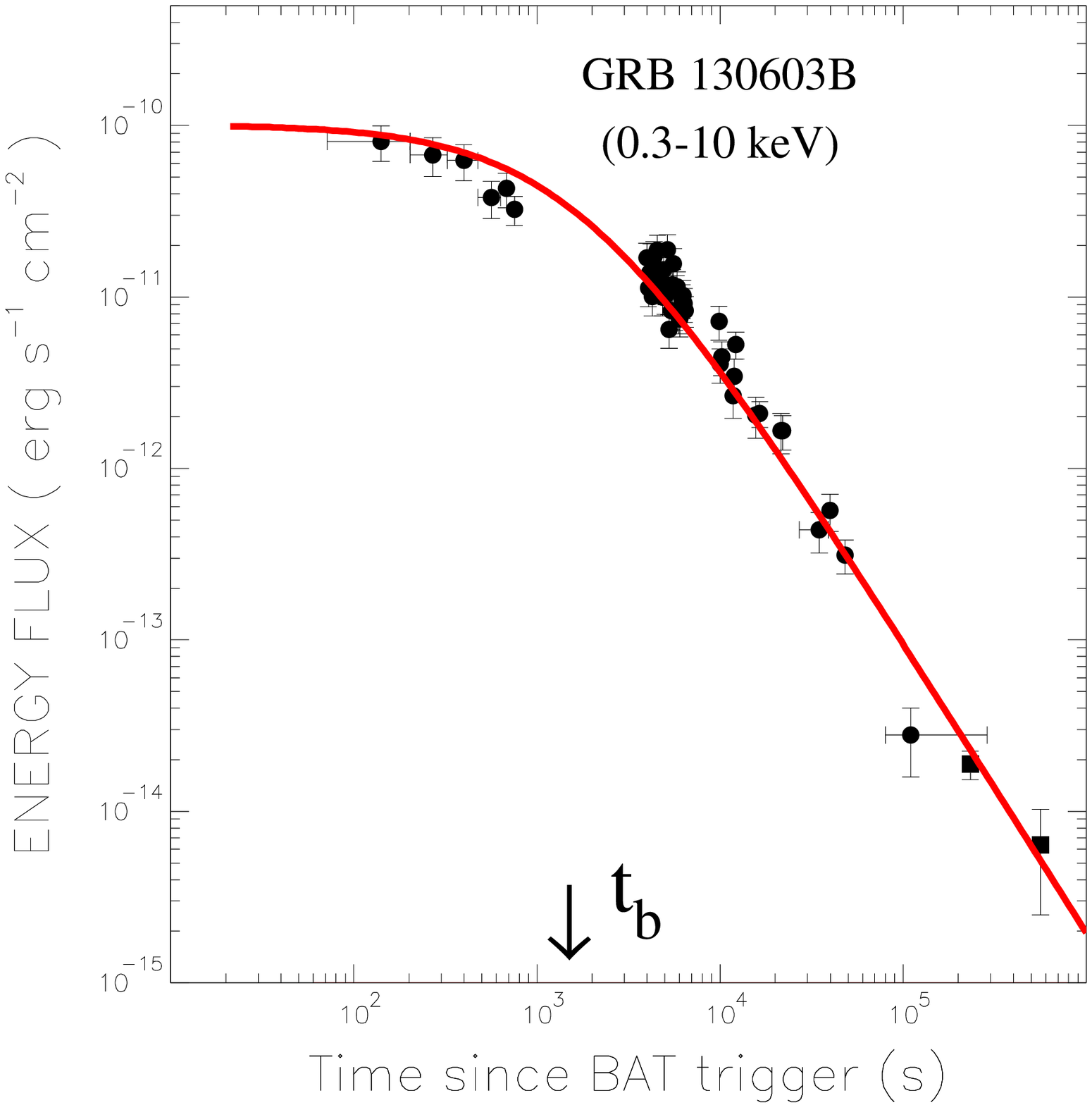,width=16.cm,height=16.cm} 
\caption{ 
Comparison between the light curve of the X-ray afterglow of 
SHB 130603B, which was measured with Swift XRT (Evans et al.~2009)
and with Newton XMM (Fong et al.~2013) assuming the spectral index 
${\rm \beta_X=1.15}$ that was measured with Swift, and a CB model fit.}
\label{fig1}
\end{figure}

\end{document}